\documentclass[reprint,aps,amsmath,amssymb,superscriptaddress,notitlepage]{revtex4-1}

\usepackage[utf8]{inputenc}
\usepackage[hidelinks]{hyperref}
\usepackage{bm,graphicx,braket,color}
\usepackage{cancel}
\usepackage{fixmath}
\usepackage{verbatim}

\begin{document}
\newcommand{\Ocal}{\mathcal{O}}
\newcommand{\Lcal}{\mathcal{L}}
\newcommand{\R}{\mathcal{R}}
\renewcommand{\vec}[1]{\mathbold{#1}}
\newcommand{\mat}[1]{#1}
\newcommand{\vphi}{\vec{\phi}}
\newcommand{\vpsi}{\vec{\psi}}
\newcommand{\vu}{\vec{u}}
\newcommand{\ve}{\vec{e}}
\newcommand{\vh}{\vec{h}}
\newcommand{\veaj}{\tilde{\vec{e}}}
\newcommand{\vhaj}{\tilde{\vec{h}}}
\newcommand{\vj}{\vec{j}}
\newcommand{\vm}{\vec{m}}
\newcommand{\vf}{\vec{f}}
\newcommand{\vF}{\vec{F}}
\newcommand{\mA}{\mat{A}}
\newcommand{\mB}{\mat{B}}
\newcommand{\vc}{\vec{c}}
\newcommand{\mJ}{\mat{J}}
\newcommand{\vg}{\vec{g}}
\newcommand{\vlam}{\vec{\lambda}}

\newcommand{\tocite}[1]{\textcolor{red}{[#1]}}
\newcommand{\param}[1]{\textcolor{black}{#1}}

\newcommand{\app}{Appendix~}




\title{Forward-mode Differentiation of Maxwell's Equations} 



\author{Tyler W. Hughes}
\author{Ian A. D. Williamson}
\author{Momchil Minkov}
\author{Shanhui Fan}
\email[]{shanhui@stanford.edu}
\affiliation{Department of Electrical Engineering and Ginzton Laboratory, Stanford University, Stanford, CA 94305, USA}

\date{\today}

\begin{abstract}
We present a previously unexplored \textit{forward-mode} differentiation method for Maxwell’s equations, with applications in the field of sensitivity analysis.  
This approach yields exact gradients and is similar to the popular adjoint variable method, but provides a significant improvement in both memory and speed scaling for problems involving several output parameters, as we analyze in the context of finite-difference time-domain (FDTD) simulations.
Furthermore, it provides an exact alternative to numerical derivative methods, based on finite-difference approximations.
To demonstrate the usefulness of the method, we perform sensitivity analysis of two problems. First we compute how the spatial near-field intensity distribution of a scatterer changes with respect to its dielectric constant.
Then, we compute how the spectral power and coupling efficiency of a surface grating coupler changes with respect to its fill factor.
\end{abstract}

\maketitle

\section{\label{sec:intro}Introduction}


The ability to differentiate Maxwell’s equations is essential to many important problems in optimization and device design.
For example, when performing inverse design of a photonic device, one typically computes the gradient of its figure of merit (FOM) with respect to numerous design parameters, which can then be used to perform optimization over a large parameter space \cite{molesky2018inverse, sigmund2003systematic, lu2013nanophotonic, hughes2017method}.
One might also wish to compute how a \textit{distribution} of output properties, such as the spatial distribution of the electromagnetic energy, changes with respect to a single design parameter, such as the material's dielectric constant. 
In general, we may consider both of these problems as instances where we wish to compute the Jacobian of a function $\vF : \R^m \mapsto \R^n$, which maps $m$ input parameters to $n$ output properties through a simulation of Maxwell's equations.


The most straightforward approach to computing the Jacobian of $\vF$ is through approximate finite-difference methods.
Here, each of the $m$ input parameters are individually perturbed by a small amount and the resulting change in outputs is measured.
As such, this technique requires at least one additional simulation per input parameter in addition to the original simulation.
On the other hand, the number of simulations scales in constant time with the number of outputs, $n$.  
This method is therefore ideal for problems with few inputs and many outputs $\left(m \ll n\right)$, such as computing the change in several properties of the system with respect to a single parameter.
In contrast, finite difference methods are highly costly in the opposing situation when $m \gg n$. 
Moreover, the derivatives computed using finite difference methods are not exact and depend crucially on the choice of numerical step size for each parameter.

The adjoint method is another common approach for computing derivatives, which may be derived using the method of Lagrange multipliers \cite{cao2003adjoint, veronis2004method, bradley_2013, hughes2018adjoint, wang2018adjoint}.
Unlike finite-difference methods, the adjoint method yields \textit{exact} gradients as it involves a numerical evaluation of the analytical Jacobian of the system.
Additionally, while it requires one additional \textit{adjoint} simulation for each output quantity, the number of simulations is constant with respect to the number of input parameters, $m$.
This makes the adjoint method optimal for many inverse design and optimization problems in photonics, where $\vF$ typically takes a large number of design parameters as the input and returns a single, scalar FOM (i.e. $n=1$ and $m \gg n$).

In this work, we introduce a third alternative method for computing the Jacobian of a function involving Maxwell's equations.
We refer to our method as \textit{forward-mode differentiation} (FMD) as it is known in the field of automatic differentiation  \cite{baydin2018automatic, rackauckas2018comparison}.
FMD is closely related to the adjoint method, which is often referred to as \textit{reverse-mode differentiation}.
Therefore, like the adjoint method, FMD provides exact gradients, but has similar time and memory scaling as finite-difference approaches.
%
%
As we will show, these properties may serve useful in a wide range of problems and FMD is, therefore, an important addition to the toolkit of numerical electromagnetics.

\begin{figure}[h]
  \centering
  \includegraphics[width=.7\columnwidth]{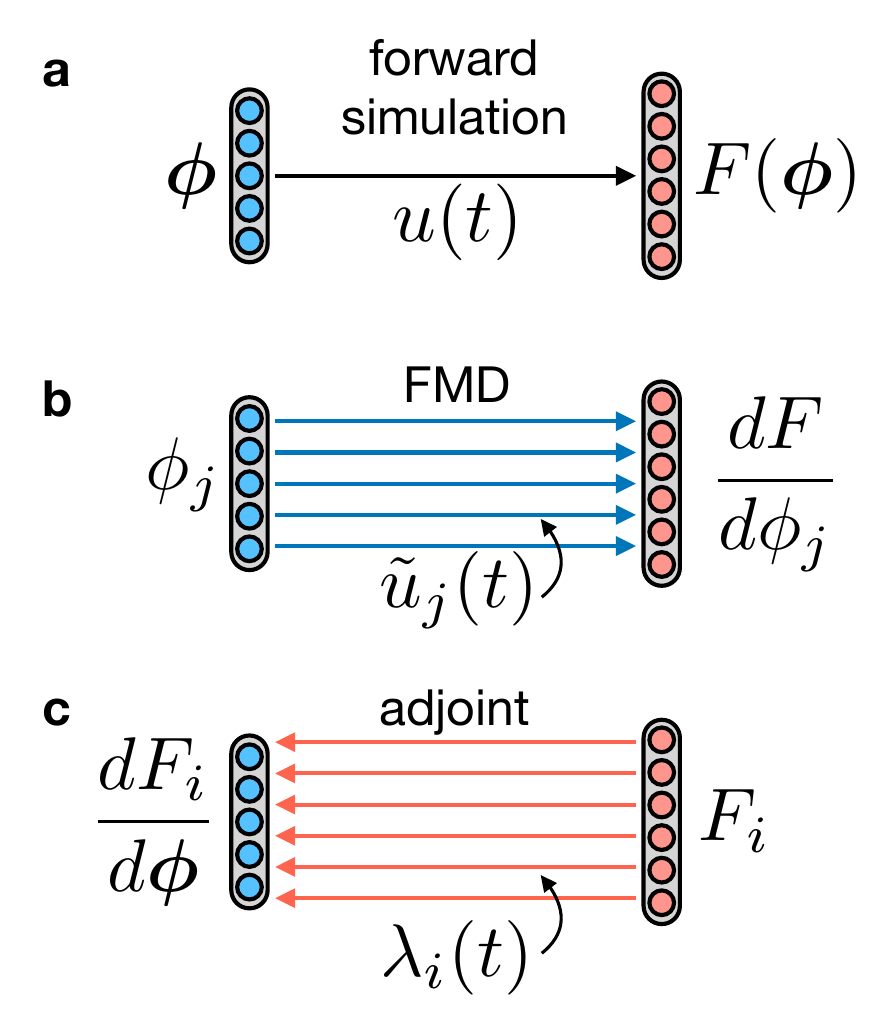}
  \caption{\textbf{Graphical comparison of adjoint and forward derivative techniques}.  \textbf{a}, The forward simulation, including the computation of $\vF(\vphi)$ from parameters $\vphi$ through an FDTD simulation with fields $\vu(t)$.  \textbf{b}, FMD requires solving one additional simulation, $\tilde{\vu}_j(t)$ for each of the $m$ parameters.  This simulation allows one to compute the change in each element of $\vF$ with respect to $\phi_j$.  \textbf{c}, The adjoint method requires solving one additional simulation, $\vlam_i(t)$ for each of the $n$ outputs.  This simulation allows one to compute the change in the $i$-th output of $\vF$ with respect to all inputs.}
  \label{fig:math}
\end{figure}

To visually compare the above three approaches, in Fig. \ref{fig:math}\textbf{a} we diagram the computation of $\vF$ as a function of input parameters $\vphi$ through a simulation of Maxwell's equations, which provide the evolution of the electromagnetic field in time as denoted by $\vu(t)$.
To compute the Jacboian of $\vF$, the FMD algorithm requires one additional electromagnetic simulation per parameter ($\tilde{\vu}_j(t)$), which we show corresponds to the forward propagation of derivative information in the system.
On the other hand, in the adjoint method, one additional electromagnetic simulation is required per \textit{output} parameter ($\vlam_i(t)$), which can be interpreted as the backward propagation of derivative information through the system \cite{insitu_2018}.

The remainder of this paper is outlined as follows.
In Section II, we first derive the basic form of FMD for a general problem involving Maxwell's equations in the time-domain and compare its mathematical structure to that of the adjoint and finite-difference methods.
In Section III, we demonstrate FMD in two practical problems: First, we analyze the derivative of an intensity pattern of a dielectric antenna with respect to the material's dielectric constant.
Then, we use FMD to analyze how the spectral coupling efficiency of a surface grating coupler changes as function of grating fill factor.
Finally, we discuss our findings in Section IV and conclude in Section V.

\section{\label{sec:derivation} Differentiation of Maxwell's Equations}

In this Section we derive the Jacobian of an electromagnetic problem using FMD and compare it to the forms given by both adjoint and finite-difference techniques.
We define our problem through a function $\vF(\vphi)$, where $\vphi \in \R^m$ is a vector of input parameters and $\vF \in \R^n$ is a vector of output properties.
For example, $\vphi$ might correspond to a set of geometric design parameters that define a photonic device and $\vF$ may be a set of performance metrics, including, for instance, the operational bandwidth or efficiency.

We assume that the evaluation of $\vF(\vphi)$ involves a solution of Maxwell's equations in the time domain, although our analysis extends to the frequency domain, for example in the context of the finite-difference frequency-domain (FDFD) method \cite{shin2012choice}, which we show in \app Section \ref{appx:fdfd}.
For concreteness, we express $\vF$ in the form
\begin{equation}
    F_i(\vphi) = \int_0^T dt~ f_i(\vu(t), t),
    \label{eq:definition}
\end{equation}
where $\vu(t) \equiv [\vh(t),~\ve(t)]^T$ is the concatenation of the magnetic and electric field vectors at time $t$.
The function, $f_i$, gives the contribution of these instantaneous field quantities to the $i$-th output property, $F_i$.
For example, if $F_i$ corresponds to the time integrated intensity at a single point in the domain, then $f_i(\vu(t), t)$ is given by $|\vu(t)|^2$ evaluated at that point.
$\vu(t)$ depends implicitly on the input parameters, $\vphi$, which define the spatial distribution of materials in the simulation domain.

The dynamics of $\vu(t)$ are governed by Maxwell's equations.
For a linear electromagnetic system with permittivity tensor $\epsilon$, permeability tensor $\mu$, electric (magnetic) conductivity tensors $\sigma_E$ ($\sigma_H$), and electric (magnetic) current sources, $\vj(t)$ ($\vm(t)$), Maxwell's equations may be written as
\begin{equation}
    \begin{bmatrix}
        \mu & 0 \\ 0 & -\epsilon
    \end{bmatrix}
    \begin{bmatrix}
        \dot{\vh}(t) \\ \dot{\ve}(t)
    \end{bmatrix}
    =
    \begin{bmatrix}
        -\sigma_H & \nabla \times \\
         \nabla \times & -\sigma_E
    \end{bmatrix}
    \begin{bmatrix}
        \vh(t) \\
        \ve(t)
    \end{bmatrix}
    +
    \begin{bmatrix}
        \vm(t) \\
        \vj(t)
    \end{bmatrix},
\label{eq:maxwell}
\end{equation}
where the spatial dependence of all of these quantities is implicit.

From here on, we assume that the input parameters, $\vphi$, only influence the $\epsilon$ and $\mu$ distributions, although the following analysis can be straightforwardly extended to other situations, such as a $\vphi$-dependent conductivity or source.
More generally, we may express Eq.~(\ref{eq:maxwell}) in terms the constraint equation
\begin{align}
  \vg(\dot{\vu}, \vu, \vphi, t) = \mA(\vphi) \cdot \dot{\vu}(t) + \mB \cdot \vu(t) + \vc(t) = \vec{0},
\label{eq:constraint}
\end{align}
where we have identified in Eq.~(\ref{eq:maxwell}) the matrices $\mA(\vphi)$ and $\mB$, as well as the source vector $\vc(t)$. 
$\vec{0}$ is defined as a vector containing all zeros.
Eq.~(\ref{eq:constraint}) is typically solved using a finite-difference time-domain (FDTD) simulation, involving discretization in the spatial and temporal domains.

We now examine three methods for computing the Jacobian of $F$, defined as $J_{ij} \equiv \frac{\partial F_i}{\partial \phi_j}$, given the constraint of Eqs. (\ref{eq:constraint}).
We will start with the finite-difference derivative approach, as it is the most simple to explain.
Then, we will introduce the FMD method and finish with the adjoint method.

\subsection{Finite-Difference Approximation}

In the finite-difference technique, one typically measures the change in the system's output given a small change in each input parameter.
Explicitly, for the $j$-th input parameter, $\phi_j$, we may approximate the derivative as a forward difference 
\begin{equation}
    \frac{d\vF}{d\phi_j} \approx \frac{\vF(\vphi + \Delta_j\vec{\hat{j}}) - \vF(\vphi)}{\Delta_j}
    \label{eq:finite_diff}
\end{equation}
where $\Delta_j$ is the numerical step size for the $j$-th parameter and $\vec{\hat{j}}$ is a vector of $0$'s except with $1$ at the $j$-th index.
Eq. (\ref{eq:finite_diff}) is evaluated by running one additional FDTD simulation for $\vF(\vphi + \Delta_j\vec{\hat{j}})$ and returns a vector specifying the derivative of each output with respect to $\phi_j$, therefore determining the $j$-th column of the Jacobian.
This operation must be performed for each element of $\vphi$, and thus the cost of computing the full Jacobian is one additional simulation per input parameter, for a total of $m$ additional simulations.

\begin{figure}[t]
  \centering
  \includegraphics[width=1\columnwidth]{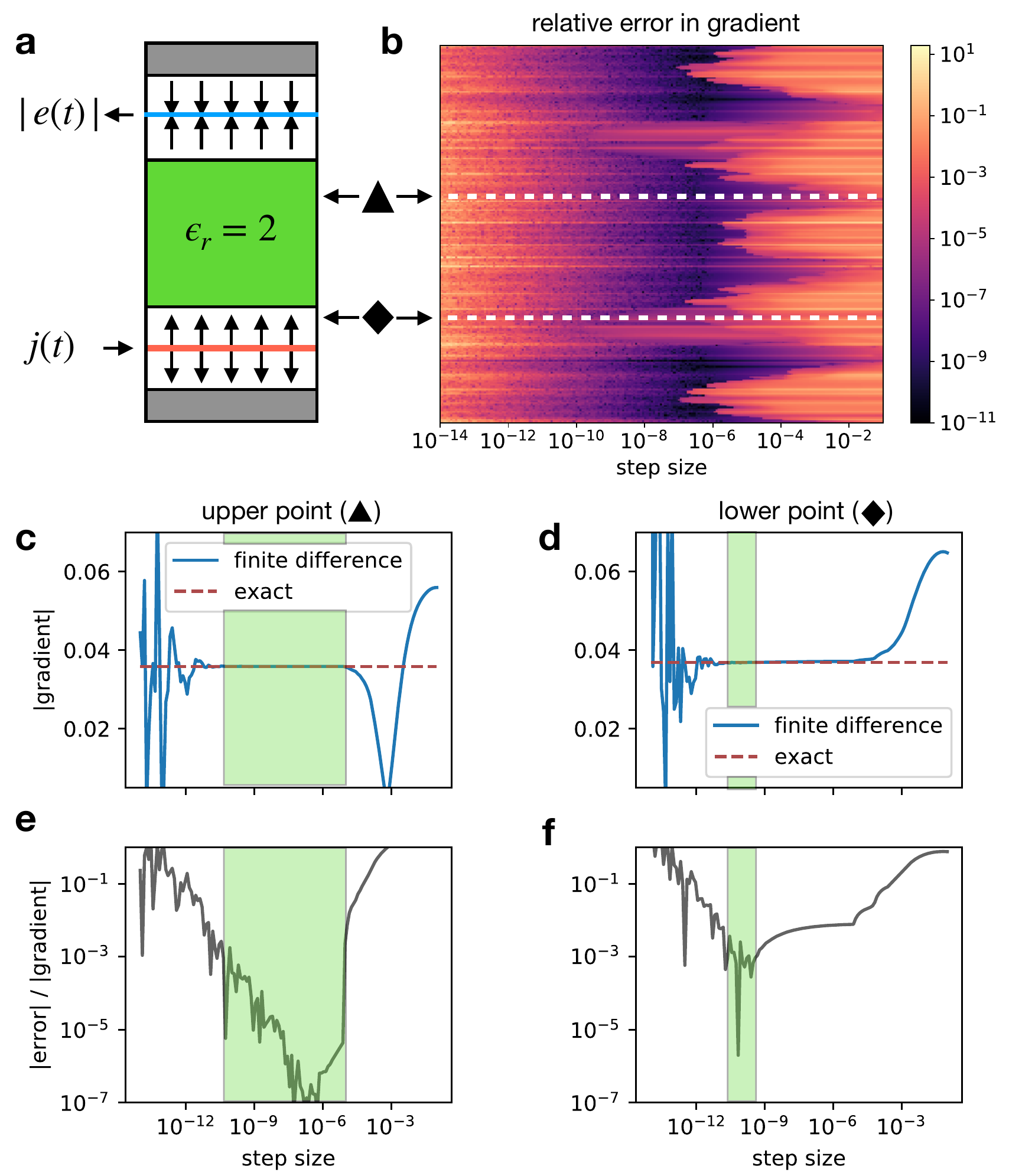}
  \caption{\textbf{Comparison of numerical and exact gradients}.
  \textbf{a}, A dielectric slab is modeled using FDTD.
  A pulse is injected into one side of the slab and the integral of $|\ve(t)|$ is measured on the other side and taken to be the figure of merit, $F$, with which the gradients are defined hereafter.
  \textbf{b}, The relative error in the gradient of $F$ with respect to the permittivity distribution as computed using numerical derivative and the exact (adjoint) derivative.
  \textbf{c,e}, For a point near the center of the slab, the accuracy of the numerical gradient is highly independent of the step size.
  \textbf{d,f}, For a point near the edge of the slab, the accuracy of the numerical gradient is highly dependent on step size.
  The Green box indicates the magnitude of the error relative to the L2 norm of the full gradient below \param{1 part in 1000}.}
  \label{fig:numerical}
\end{figure}

In practice, this finite differences are not ideal as they yield only approximate derivatives and require the determination of a step size, $\Delta_j$, for each parameter.
To explore this issue, in Fig. \ref{fig:numerical} we examine the accuracy of the finite-difference derivative as it compares to an exact method, such as FMD or the adjoint method. 
We define a problem corresponding to transmission through a dielectric slab, as diagrammed in  Fig. \ref{fig:numerical}a.
The domain is one-dimensional with perfectly-matched layers on the vertical boundaries to absorb outgoing waves.
We inject a pulse into one side of the slab and measure the integral of the absolute value of the electric field at a probe on the other side of the slab, namely $F(\vphi) = \int dt~ |\vec{p}^T \ve(t)|$, where $\vec{p}$ defines the probe location.

We then compute the gradient of $F$ with respect to the permittivity of each grid cell within the domain using a finite-difference approach, as in Eq. \ref{eq:finite_diff}, and the adjoint method, which is exact.
The relative error in the gradient is plotted in Fig. \ref{fig:numerical}b.
The $y$ axis of this plot corresponds to the diagram in Fig. \ref{fig:numerical}a.
We observe that points near the boundary of the slab are highly sensitive to the numerical step size parameter.

In Fig. \ref{fig:numerical}c-f we inspect the relative errors with respect to the permittivities at two distinct points, as annotated by the diamond and triangle shapes in \ref{fig:numerical}b.
As seen in \ref{fig:numerical}c-d, the required numerical step sizes that leads to low error (indicated by the green box) are vastly different for these two points.
These findings suggest difficulties of applying the finite difference method for gradient calculations in general.
These difficulties serve as an issue not only for practical applications in sensitivity analysis, but also when using numerical derivatives as a point of comparison when confirming the correctness of implementations of exact methods.

\subsection{Forward-Mode Differentiation}

We now introduce the FMD method, which is the focus of this work.
To derive this, we first directly differentiate the figure of merit $\vF$ from Eq. (\ref{eq:definition}) with respect to the $j$-th parameter, $\phi_j$, which gives
\begin{equation}
    \frac{d \vF}{d \phi_j} = \int_0^T dt~ \frac{\partial \vf}{\partial \vu}(t) \cdot  \frac{d\vu}{d\phi_j}(t) \\
    \label{eq:forward_diff_start}
\end{equation}
The form of the matrix $\frac{\partial \vf}{\partial \vu}(t)$ may be solved analytically and evaluated numerically using the solution of $\vu(t)$.
To evaluate $\frac{d\vu}{d\phi_j}(t)$, we differentiate the constraint equation $\vg(\dot{\vu}, \vu, \vphi, t)$ of Eq. (\ref{eq:constraint}) with respect to $\phi_j$.
This derivative gives the following expression
\begin{align}
      & \frac{d}{d \phi_j}\vg(\dot{\vu}, \vu, \vphi, t) = \vec{0}\\
      =& \frac{\partial \vg}{\partial \dot{\vu}} \cdot \frac{d\dot{\vu}}{d\phi_j}
       + \frac{\partial \vg}{\partial \vu} \cdot \frac{d\vu}{d\phi_j}
       + \frac{\partial \vg}{\partial \phi_j}
    \label{eq:forward_constraint},
\end{align}
which now may be interpreted as a new constraint for the quantity $\frac{d\vu}{d\phi_j}$.  

Like the original constraint equation, Eq. (\ref{eq:forward_constraint}) may be expressed in the form of Maxwell's equations.
To show this, we define $\frac{d\vu}{d\phi_j}(t) \equiv [\vh_j(t), \ve_j(t)]^T$ as the `derivative' fields for parameter $\phi_j$ and evaluate the other terms using (\ref{eq:maxwell}) and (\ref{eq:constraint}), giving
\begin{equation}
     ~\mA(\vphi) \frac{d\dot{\vu}}{d\phi_j}
       + \mB \frac{d\vu}{d\phi_j}
       + \frac{\partial \mA}{\partial \phi_j}\cdot \dot{\vu} =  \vec{0} \\
\end{equation}
or in the form of Maxwell's equations,
\begin{equation}
    \begin{bmatrix}
        \mu & 0 \\ 0 & -\epsilon
    \end{bmatrix}
    \begin{bmatrix}
        \dot{\vh_j} \\ \dot{\ve_j}
    \end{bmatrix}
    =
    \begin{bmatrix}
        -\sigma_H & \nabla \times \\
         \nabla \times & -\sigma_E
    \end{bmatrix}
    \begin{bmatrix}
        \vh_j \\
        \ve_j
    \end{bmatrix}
    +
    \begin{bmatrix}
        -\frac{\partial \mu}{\partial \phi_j} \cdot \dot{\vh} \\
        \frac{\partial \epsilon}{\partial \phi_j} \cdot \dot{\ve}
    \end{bmatrix}.
\label{eq:constraint2}
\end{equation}
Interestingly, from Eq. (\ref{eq:constraint2}), we notice that the derivative fields $\vh_j$ and $\ve_j$ evolve according to a similar Maxwell's equation as the original fields of Eq. (\ref{eq:maxwell}).
However, the source term is now replaced by $[-\frac{\partial \mu}{\partial \phi_j} \cdot \dot{\vh},\frac{\partial \epsilon}{\partial \phi_j} \cdot \dot{\ve}]^T$, where $\frac{\partial \epsilon}{\partial \phi_j}$ ($\frac{\partial \mu}{\partial \phi_j}$) is the change in the permittivity (permeability) with respect to the $j$-th input parameter.
We note that this source term depends explicitly on the fields from the original simulation ($\dot{\ve}$ and $\dot{\vh}$).

With the quantity $\frac{d\vu}{d\phi_j}(t)$ solved by running an additional FDTD simulation as defined by Eq. (\ref{eq:constraint2}), one may plug this into Eq. (\ref{eq:forward_diff_start}) to evaluate the $j$-th column of the Jacobian. 
Therefore, like the finite-difference method, one must repeat this process with a new FDTD simulation for each of the $m$ input parameters to compute the full Jacobian.
However, unlike the finite-difference method, the gradients evaluated here are $\textit{exact}$.

\subsection{Adjoint Method}

For contrast, we now briefly describe the adjoint method, with a full derivation included in the \app Section \ref{appx:adjoint}.
In the adjoint method, one is interested in computing the same quantity as the previous section, namely the Jacobian of $\vF$ with respect to $\vphi$.
While the adjoint and FMD methods are mathematically equivalent, they evaluate the Jacobian in reverse order from each other.
Whereas in FMD, the derivative simulation is forward propagated through the system for each input parameter, in the adjoint method, one propagates an `adjoint' simulation backwards through the system for each output parameter.

One first solves the `forward' problem, corresponding to running an FDTD simulation for $\vu(t)$.
Then, one must solve a second `adjoint' problem for the $i$-th output parameter, which defines the solution $\vlam_i(t)~\equiv~[\vhaj_i(t),~\veaj_i(t)]^T$, governed by the following constraint
\begin{equation}
    \frac{\partial \vg}{\partial \dot{\vu}}^T \cdot \dot{\vlam}_i
    - \left( \frac{\partial \vg}{\partial \vu}^T - \frac{d}{dt}\frac{\partial \vg}{\partial \dot{\vu}}^T \right) \cdot \vlam_i
    - \frac{\partial f_i}{\partial \vu}^T = \vec{0}.
    \label{eq:adjoint}
\end{equation}
Crucially, the adjoint solution has a boundary condition of $\vlam_i(T) = \vec{0}$.
This means that it must be solved backwards in time from $t=T$ to $t=0$.
We also note that, like the FMD method, its source term $\frac{\partial f_i}{\partial \vu}^T$ depends explicitly on the forward solution, $\vu(t)$.

Expressing the adjoint constraint in terms of Maxwell's equations gives the following electromagnetic simulation
\begin{equation}
 \mA^T(\vphi) \dot{\vlam}_i - \mB^T \vlam_i - \frac{\partial f_i}{\partial \vu}^T = \vec{0}\\
\end{equation}
or in terms of Maxwell's equations
\begin{equation}
    \begin{bmatrix}
        -\mu^T & 0 \\ 0 & \epsilon^T
    \end{bmatrix}
    \cdot
    \begin{bmatrix}
        \dot{\vhaj}_i \\ \dot{\veaj}_i
    \end{bmatrix}
    =
    \begin{bmatrix}
        -\sigma_H^T & \nabla \times \\
         \nabla \times & -\sigma_E^T
    \end{bmatrix}
    \cdot
    \begin{bmatrix}
        \vhaj_i \\ \veaj_i
    \end{bmatrix}
    +
    \begin{bmatrix}
        \frac{\partial f_i}{\partial \vh} ^T \\
        \frac{\partial f_i}{\partial \ve} ^T
    \end{bmatrix}.
    \label{eq:adjoint_matrix}
\end{equation}
Interestingly, the evolution of the adjoint fields is the same as the forward fields if all of the following substitutions are made
\begin{enumerate}
\item{$t \to T - t$, which corresponds to time reversal and subsequent shifting by the total simulation time $T$ to match boundary conditions at $t=T$.}
\item{$\epsilon \to \epsilon^T$, $\mu \to \mu^T$, $\sigma_E \to \sigma_E^T$, and $\sigma_H \to \sigma_H^T$.  If the usual case that the system obeys Lorentz reciprocity, then these quantities are symmetric, in which case the adjoint system is the same as the original system.}
\item{$\vm(t) \to \frac{\partial f_i}{\partial \vh}(T-t)^T$ and $\vj(t) \to \frac{\partial f_i}{\partial \ve}(T-t)^T$ which corresponds to setting a new source for the adjoint fields that depends on the figure of merit's dependence on the solution $\vu(t)$.}
\end{enumerate}

With the adjoint solution $\vlam_i(t)$ found, one may compute the gradient of $F_i$ now as 
\begin{align}
    \frac{dF_i}{d\vphi} &= \int_0^T dt~ \vlam_i(t)^T \cdot \frac{\partial \vg}{\partial \vphi}(t)\\
    &= \int_0^T dt~ \vlam_i(t)^T \cdot \frac{\partial \mA}{\partial \vphi} \cdot \vu(t)
    \label{eq:final_grad_adj},
\end{align}
where $\frac{\partial \mA}{\partial \vphi}$ is a rank 3 tensor.
As Eq. (\ref{eq:final_grad_adj}) returns the $i$-th row of the Jacobian, to compute the full Jacobian, one must run an additional adjoint simulation for each output parameter.
This is in contrast with the finite difference and FMD methods, which need one additional simulation per \textit{input} parameter.
In practice, a more efficient method for computing Eq. (\ref{eq:final_grad_adj}) can be performed, which is outlined in \app Section \ref{appx:efficient}.

A full comparison of the time and memory complexity of the two methods is summarized in Table \ref{tb:table_order}, with a detailed explanation in \app Section \ref{appx:scaling}.

\begin{table}[]
\begin{tabular}{|l|l|l|}
\hline
\textbf{Method} & \textbf{Time Complexity}  & \textbf{Memory Complexity}   \\ \hline
Finite Difference     & $\Ocal(NTm)$   & $\Ocal(N)$ \\ \hline
FMD & $\Ocal(NTm)$ & $\Ocal(NT+Nn)$ \\ \hline
Adjoint & $\Ocal(NTn)$ & $\Ocal(NT+Nm)$  \\ \hline
\end{tabular}
\caption{Comparison of memory and speed complexity of the three different gradient computation methods examined in this work.  $N$ is the number of grid cells.  $T$ is the number of time steps. $m$ and $n$ are the number of input and output variables, respectively, of the function $F$.  We assume $n$ and $m$ are each less than or equal to $N$.}
\label{tb:table_order}
\end{table}

\section{Demonstrations}

To demonstrate the FMD method, we now apply it to two sample problems.  First, we will show that one can use FMD to compute the exact derivative of a spatial distribution, in this case, the electric field intensity distribution, with respect to design parameters.  Then, we show that one can use FMD to compute the derivative of outputs as a function of frequency, using a grating coupler as an example.

\subsection{Intensity Distribution of a Scatterer}

\begin{figure}[t]
  \centering
  \includegraphics[width=1\columnwidth]{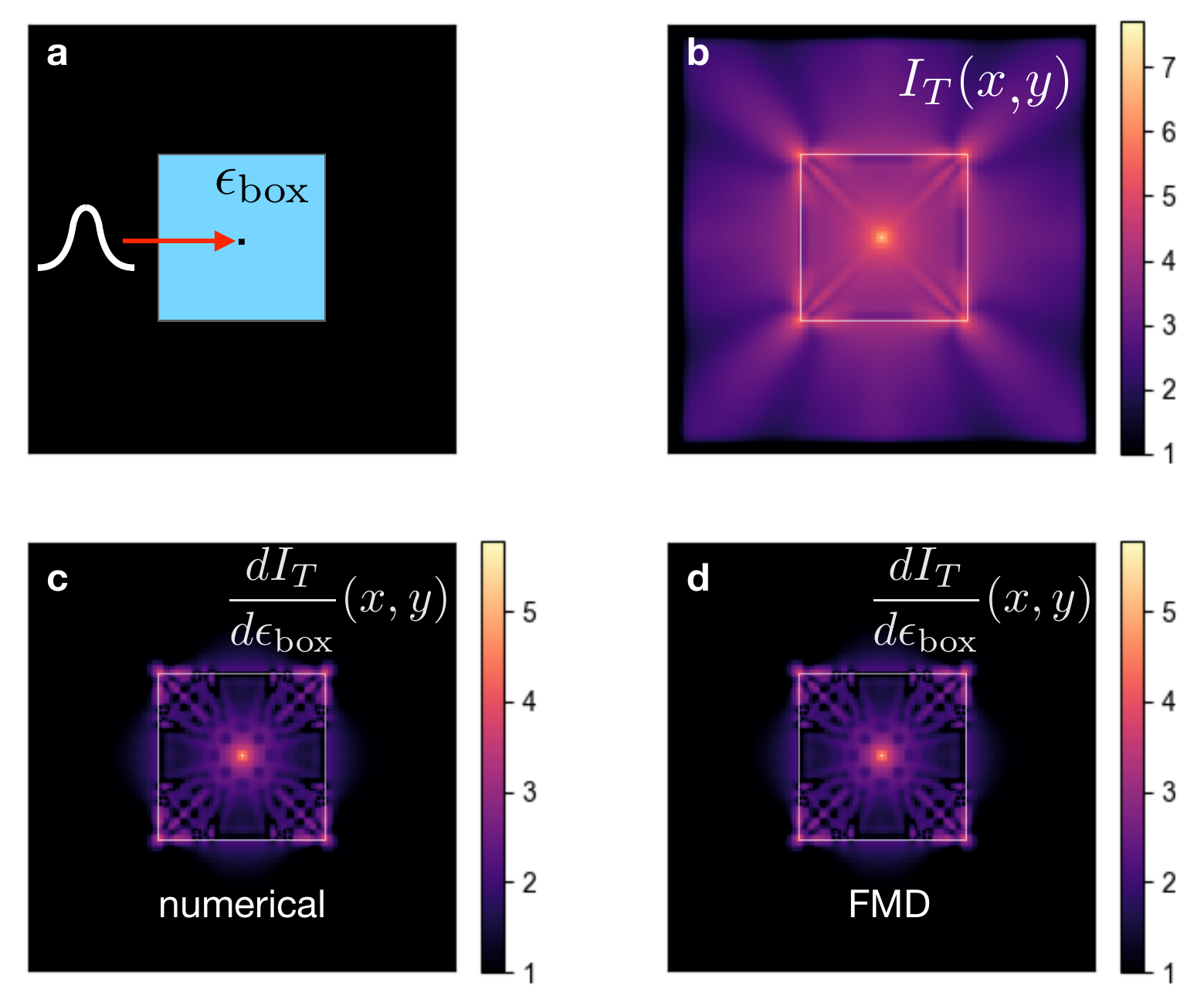}
  \caption{\textbf{Comparison of FMD and numerical derivative on sample problem}. \textbf{a}, Problem setup.  A Gaussian pulse is injected into the center of a dielectric square with relative permittivity \param{$\epsilon_\textrm{box} = 12$}.  \textbf{b}, The logarithm of the resulting time-integrated intensity distribution $I_T(x,y) = \int\int dx~dy~I(x,y,t)$ is shown at each point in space. \textbf{c}, Using an approximate numerical derivative with step size of $10^{-3}$, we plot the logarithm of the change in $I_T(x,y)$ with respect to the dielectric function of the box, requiring one simulation. \textbf{d}, using exact FMD, we plot the same derivative.}
  \label{fig:demo}
\end{figure}

In Fig. \ref{fig:demo}\textbf{a}, we simulate a two-dimensional domain with a point emitter located at the center of a dielectric square with permittivity $\epsilon_\textrm{box}$ and length of \param{410 nm}.
The surrounding medium is vacuum with \param{10} grid cells of perfectly-matched absorbing layers (PMLs) on each side.
We inject a DC pulse with a temporal width of \param{289 fs} into the box and measure the optical intensity distribution at each grid cell, integrated over time $I_T(x,y) \equiv \int_0^T ~dt~I(x,y,t)$, which is shown in Fig. \ref{fig:demo}\textbf{b}.
\textbf{}
Then, we compute the derivative of this intensity distribution with respect to the permittivity of the box in two ways.
First, we compute this using a numerical derivative, where $\epsilon_\textrm{box}$ is increased and decreased by \param{$1\times 10^{-3}$} and a derivative is approximated using central finite difference.
The result is shown in Fig. \ref{fig:demo}\textbf{c}.
Then, we compute the same derivative in an \textit{exact} form using FMD, which requires only one additional FDTD simulation.
The resulting sensitivity pattern is shown in Fig. \ref{fig:demo}\textbf{d} and agrees with the numerical result to good precision.

\subsection{Grating Coupler Efficiency Spectrum}

\begin{figure}[t]
  \centering
  \includegraphics[width=1\columnwidth]{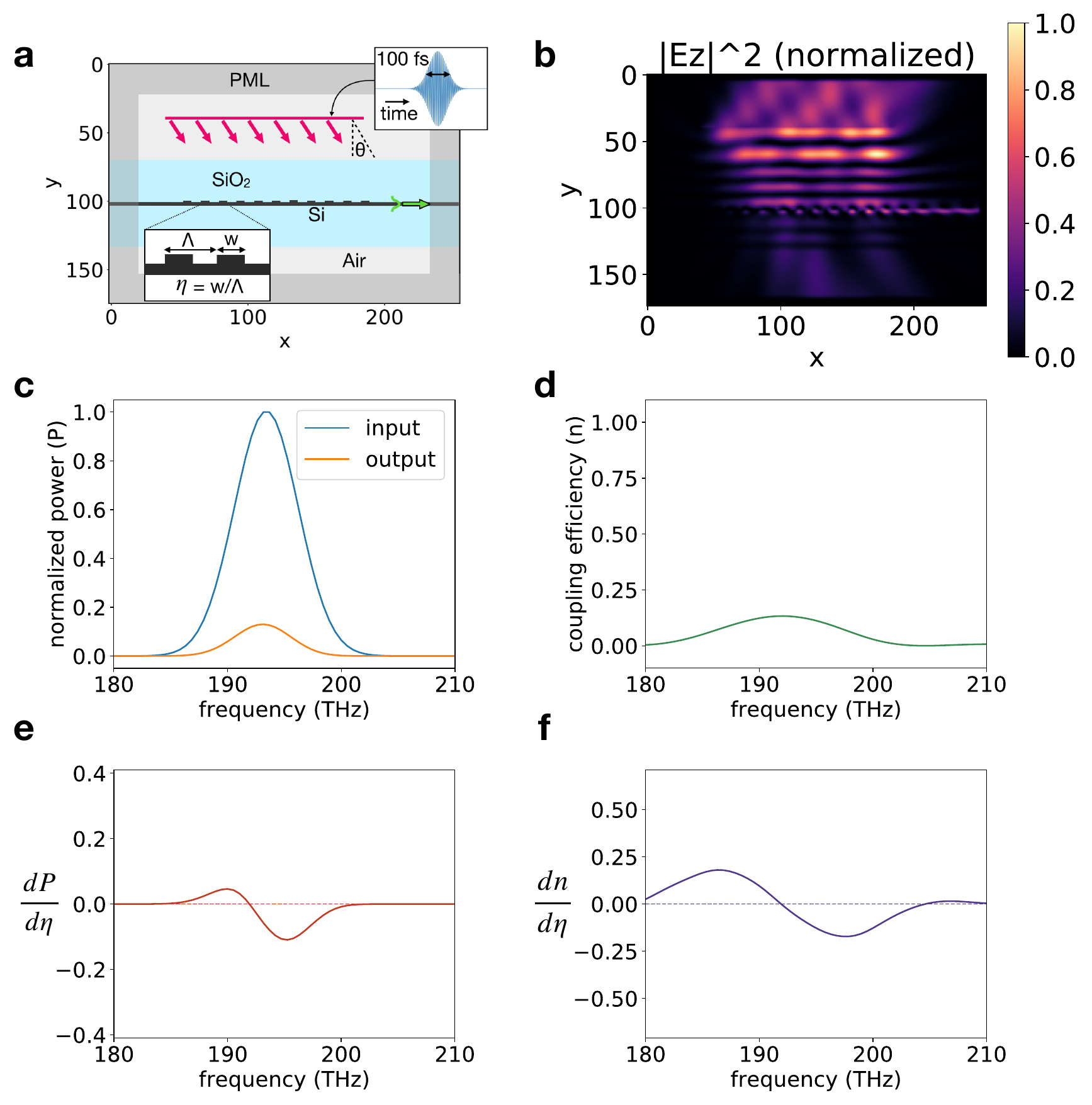}
  \caption{\textbf{FMD analysis of the spectrum of a grating coupler}. \textbf{a}, A Si grating coupler (gray) is encased in a SiO$_2$ substrate (blue).  A line source (red) above emits a pulse centered at \param{$\lambda_0 = 1550$ nm} with FWHM \param{100 fs} at angle \param{$\theta$ = 20} degrees.  The power in the waveguide mode (green) is measured as a function of frequency.  \textbf{b}, Frequency domain simulation of the out of plane electric field ($|E_z|^2$) normalized to its maximum value at the central frequency, showing good coupling to the waveguide mode. \textbf{c}, the normalized input power (blue) and the normalized power measured in the waveguide (orange) as a function of frequency. \textbf{d},  The coupling efficiency as a function of frequency ($n$).  \textbf{e}, The derivative of the coupled power with respect to the fill factor ($\eta$) of the grating, computed with FMD as a function of frequency. \textbf{f}, The derivative of the coupling efficiency ($n$) with respect to the fill factor ($\eta$) of the grating, computed with FMD as a function of frequency.}
  \label{fig:demo_grating}
\end{figure}

Now, we show that FMD is a useful tool for performing \textit{spectral} sensitivity analysis, using a grating coupler as an example \cite{su2018fully, sapra2019inverse}.
Like the previous example, wherein the gradient was taken over the entire spatial domain, this is another problem where the differentiation is needed for several outputs, in this case over the frequency domain.
In Fig. \ref{fig:demo_grating}\textbf{a}, we outline a typical grating coupler setup where a free space Gaussian pulse is coupled into a guided mode through a surface grating.
We wish to compute how the coupling performance depends on the fill-factor ($\eta$) of the grating, defined as the ratio of the grating width to the grating period.
We inject a pulse centered at \param{$\lambda_0$ = 1550 nm} with a duration of \param{$100$ fs} into the top of the domain via a finite-width line source emitting at an angle \param{$\theta = 20$} degrees from normal incidence.
A Si grating structure is encased in a SiO$_2$ substrate of thickness \param{1 $\mu$m} on each side.
The base thickness of the coupler is \param{150 nm} and the tooth height is \param{70 nm}, corresponding to an etched SOI platform with a \param{220 nm} thick Si layer.
For a fill factor of 0.5, an optimal grating period of \param{660 nm} is computed using the effective index of the grating structure, the free space pulse wavelength, and the incident angle, following Ref. \cite{chrostowski2015silicon}.
A PML of \param{10} grid cells is included on all edges of the domain for absorbing boundary conditions.
The structure is simulated using FDTD and the power in the waveguide mode is measured as a function of frequency.

In Fig. \ref{fig:demo_grating}\textbf{b}, we show a frequency domain simulation of the structure at \param{$\lambda_0 = 1550$ nm}, showing good coupling between the incident light and the waveguide mode.
In Fig. \ref{fig:demo_grating}\textbf{c}, we plot the incident power spectrum normalized by its maximum value, using a time domain simulation.
This is compared to the power measured in the waveguide mode, normalized by the same value.
By comparing the integrals of these curves over the full frequency range, we compute a total coupling efficiency of \param{11.3\%}.
Fig. \ref{fig:demo_grating}\textbf{d} shows the coupling efficiency of the device as a function of input frequency.
In Fig. \ref{fig:demo_grating}\textbf{e} we show the derivative of the coupled power (corresponding to orange curve in Fig. \ref{fig:demo_grating}\textbf{c}) with respect to the fill factor of the grating, using FMD.
In Fig. \ref{fig:demo_grating}\textbf{f} we show the derivative of the coupling efficiency (corresponding to Fig. \ref{fig:demo_grating}\textbf{d}) with respect to the fill factor of the grating, using FMD.

This results here provide another demonstration of how FMD may be used to compute the \textit{exact} derivative of a cost function with multiple components with respect a single input parameter using just one additional simulation.

\section{\label{sec:disc}Discussion}

In this work, we introduced a forward differentiation method for computing the gradient of a figure of merit that is a function of an electromagnetic FDTD simulation.  We have shown that this method serves as an attractive alternative to both adjoint-based and numerical gradient calculation methods.
For problems where there are more output parameters than input parameters, the benefits of this method over the adjoint method are significant.
Furthermore, this approach eliminates the need to determine a numerical step size for each parameter, the optimal value of which is generally difficult to determine with additional simulations.

Whereas forward differentiation is an approach that is mentioned in applied math literature in the subject of automatic differentiation, to our knowledge, it has never been directly applied to an electromagnetic simulation.
An approach known as `complex step differentiation' had previously been applied to FDTD \cite{sarris2015broadband}.
In this approach, an imaginary-valued perturbation to each parameter is applied, and the resulting finite-difference derivative suffers from far less numerical error.
While this technique shares many of the benefits of forward differentiation, it also requires a numerical step size and additional complications, including a mechanism for handling complex-valued electromagnetic fields in FDTD.
In the quantum information processing community, a similar approach has been proposed for measuring exact gradients through forward propagation of error signals \cite{schuld2019evaluating}.
While this is an interesting technique that has parallels to FMD, it requires specific conditions on the mathematical form of the system which are not required in FMD.

As mentioned, the FMD approach is not preferred when considering inverse design problems with few design objectives and multiple degrees of freedom in the design parameters.
For these applications, an adjoint method is highly preferred in terms of speed.
However, there are many instances where FMD may be a useful complement to adjoint methods.
For example, one may use FMD to compute the sensitivity of a device's performance when a dilation or contraction is applied to its geometric distribution \cite{wang2011robust, wang2019robust}.
Additionally, the simplicity of the FMD method makes it a good alternative to the adjoint method for inverse design parameters involving few parameters, such as photonic crystal optimization \cite{minkov2014automated}.
Finally, FMD is a better alternative to finite-difference derivatives when verifying the correctness of implementations of more complicated, exact methods, such as the adjoint method.

To make the FMD and adjoint method presented in this work more accessible, we have released an open-source FDTD and FDFD package that features gradients computed by all three methods outlined in this paper \cite{ceviche_2019}.
Our implementation makes use of automatic differentiation to provide flexible usage and more robust computation \cite{paszke2017automatic, laporte2019highly, hughes2019wave} 

\section{\label{sec:conc}Conclusion}

In conclusion, we have introduced a `forward differentiation' method for computing derivatives of quantities computed via electromagnetic simulations.
This method may be thought of as an `exact' alternative to numerical, finite-difference approaches to derivative computation computation.
Furthermore, we have shown that this method is preferable to the adjoint method, in terms of time complexity, for problems involving more output quantities than input parameters.
As such, this work will present a useful alternative to existing gradient computation methods and enable more efficient modeling and design of a wide range of components that are modelled using Maxwell's equations.

\section{Acknowledgements}

This work is supported by the Gordon and Betty Moore Foundation (GBMF4744); the Swiss National Science Foundation (P300P2\_177721); and the Air Force Office of Scientific Research (AFOSR) (FA9550-17-1-0002, FA9550-18-1-0379).

\bibliography{bib}


\onecolumngrid
\pagebreak
\widetext
\newpage 
\begin{center}
\section*{\textbf{\large \app}}
\end{center}
\setcounter{equation}{0}
\setcounter{figure}{0}
\setcounter{table}{0}
\setcounter{section}{0}
\setcounter{page}{1}
\makeatletter
\renewcommand{\thepage}{S\arabic{page}}
\renewcommand{\thetable}{S\arabic{table}}
\renewcommand{\theequation}{S\arabic{equation}}
\renewcommand{\thefigure}{S\arabic{figure}}

\section{Frequency-Domain Forward-Mode Differentiation \label{appx:fdfd}}

In a non-magnetic material with $\mu = \mu_0$, Maxwell's equations at steady state when driven by frequency $\omega$ are described as 
\begin{equation}
    \nabla \times \nabla \times \ve - \left(\frac{\omega}{c_0}\right)^2 \epsilon \ve = i \omega \vj,
    \label{eq:FDFD_1}
\end{equation}
where $\ve$ is now a phasor and the physical electric fields have time dependence $\R \left\{ \ve~e^{i \omega t} \right\}$.
The magnetic fields $\vh$ may be found by applying Maxwell's equations to the electric field solution $\ve$.

Eq. (\ref{eq:FDFD_1}) is typically written in the more general form
\begin{equation}
    A(\epsilon) \ve = \vec{b},
\end{equation}
which is solved for $\ve$ by finding
\begin{equation}
    \ve = A(\epsilon)^{-1} \vec{b},
\end{equation}

As in the time domain case, let us now consider a function $\vF(\vphi)$ where the $i$-th element is computed from the electric field distribution through $F_i = f_i(\ve(t))$.
We now compare the computation of $\frac{d\vF}{d\vphi}$ through the adjoint and FMD methods.
To do this, following a treatment given in \cite{hughes2018adjoint}, we apply the chain rule to obtain
\begin{align}
    \frac{d\vF}{d\vphi} &= \frac{\partial \vf}{\partial \ve} \cdot
                          \frac{d \ve}{d \vphi} + \frac{\partial \vf}{\partial \ve^*} \cdot \frac{d \ve^*}{d \vphi} \\
                        &= 2 \R \left\{ \frac{\partial \vf}{\partial \ve} \cdot
                          \frac{d \ve}{d \vphi} \right\} \\
                        &= -2 \R \left\{ \frac{\partial \vf}{\partial \ve} \cdot
                          A^{-1} \frac{\partial A}{\partial \vphi} \ve \right\}
                          \label{eq:deriv_fdfd}
\end{align}

In the adjoint method, Eq. (\ref{eq:deriv_fdfd}) is solved by applying a transpose and then evaluating the adjoint field
\begin{equation}
    \ve_\textrm{adj} = -A^{-T} \frac{\partial \vf}{\partial \ve}^T
\end{equation}
before plugging in the result to obtain
\begin{equation}
    \frac{d\vF}{d\vphi}^{(\textrm{adj})} = 2 \R \left\{ \ve_\textrm{adj}^T \frac{\partial A}{\partial \vphi} \ve \right\}
\end{equation}

In FMD, however, we directly apply $A^{-1}$ to the right of this expression, solving for the FMD derivative field 
\begin{equation}
    \ve_\textrm{FMD} = A^{-1} \frac{\partial A}{\partial \vphi} \ve
\end{equation}
and expressing the final result as 
\begin{equation}
    \frac{d\vF}{d\vphi}^{(\textrm{FMD})} = -2 \R \left\{ \frac{\partial \vf}{\partial \ve} \cdot \ve_\textrm{FMD} \right\}
\end{equation}

The correspondence with the time domain formalism of FMD given in the main text is apparent by inspetion.

\section{Derivation of Adjoint Sensitivity for FDTD \label{appx:adjoint}}

In this Section, we derive the form of the adjoint sensitivity that was used in Section \ref{sec:derivation} of the main text.
As before, we wish to compute the Jacobian of function $\vf$ with respect to its inputs $\vphi$, defined as follows
\begin{equation}
    \vF(\vphi) = \int_0^T dt~ \vf(\vu(t), t) \\
    \label{eq:forward_diff_start_aj}
\end{equation}
where the vector $\vu(t)$ is given by the solution to an FDTD simulation, defined as the constraint.
\begin{equation}
    \vg(\dot{\vu}, u, \vphi, t) = A(\vphi) \cdot \dot{\vu}(t) + B \cdot \vu(t) + \vc(t) = 0.
\end{equation}
For convenience we set the initial condition $\dot{\vu}(0) = \vu(0) = 0$.
As shown in the main text, the adjoint method must be applied once to compute the derivative of each output of $\vF$.
Therefore, for convenience of notation, we consider a scalar function $F$, with the understanding that the same method may be applied to each output of a vector function $\vF$.

Here we derive the gradient of $F$ using the adjoint method by an application of Lagrange multipliers.
One first defines the Lagrangian
\begin{equation}
    \Lcal = \int_0^T dt~ \left[ f(\vu(t, \vphi), t) + \vlam(t)^T \vg(\dot{\vu}, \vu, \vphi, t) \right],
    \label{eq:lagrangian}
\end{equation}
where $\lambda(t)$ is a vector of Lagrange multipliers.
We note that, when the constraints are satisfied, $\vg= 0$, which means $\Lcal = F$ and $\frac{d\Lcal}{d\vphi} = \frac{dF}{d\vphi}$.  
Differentiating Eq. (\ref{eq:lagrangian}) with respect to $\vphi$ gives the following

\begin{equation}
    \frac{d\Lcal}{d\vphi} = \int_0^T dt~ \left[ \frac{\partial f}{\partial \vu} \frac{d\vu}{d\vphi} 
    + \vlam^T \frac{\partial \vg}{\partial \dot{\vu}} \frac{d\dot{\vu}}{d\vphi}
    + \vlam^T \frac{\partial \vg}{\partial \vu} \frac{d\vu}{d\vphi} + \vlam^T \frac{\partial \vg}{\partial\vphi} \right]
    \label{eq:lagrangian_der1}
\end{equation}

The second term may be integrated by parts as follows
\begin{equation}
    \int_0^T dt~ \vlam^T \frac{\partial \vg}{\partial \dot{\vu}} \frac{d\dot{\vu}}{d\vphi} 
    = \vlam^T \frac{\partial \vg}{\partial \dot{\vu}} \frac{d\vu}{d\vphi} \Big|^T_0 - \int_0^T dt~ \left[ \dot{\vlam}^T \frac{\partial \vg}{\partial \dot{\vu}} + \vlam^T \frac{d}{dt} \frac{\partial \vg}{\partial \dot{\vu}} \right] \frac{d\vu}{d\vphi},
\end{equation}
which, when reinserted into Eq. (\ref{eq:lagrangian_der1}), gives

\begin{equation}
    \frac{d\Lcal}{d\vphi} = \int_0^T dt~  \Bigg[ \left( \frac{\partial f}{\partial \vu}
    - \dot{\vlam}^T \frac{\partial \vg}{\partial \dot{\vu}}
    + \vlam^T \left[\frac{\partial \vg}{\partial \vu} - \frac{d}{dt} \frac{\partial \vg}{\partial \dot{\vu}}
    \right] \right) \frac{d\vu}{d\vphi} + \vlam^T \frac{\partial \vg}{\partial \vphi} \Bigg]
    + \vlam^T \frac{\partial \vg}{\partial \dot{\vu}} \frac{d\vu}{d\vphi} \Big|^T_0
    \label{eq:lagrange_deriv2}
\end{equation}

We now wish to choose $\vlam$ to eliminate terms of $\frac{du}{d\phi}$ from our expression, as these terms are not directly computable.  First, we choose the condition $\vlam(T) = \vec{0}$, which, along with our original condition of $\vu(0) = \vec{0}$, eliminates the term outside of the integral.  To handle the terms within the integral, we may express $\vlam(t)$ as the solution to

\begin{equation}
    \dot{\vlam}^T \frac{\partial \vg}{\partial \dot{\vu}} - \vlam^T \left( \frac{\partial \vg}{\partial \vu} - \frac{d}{dt}\frac{\partial \vg}{\partial \dot{\vu}}\right)  - \frac{\partial f}{\partial \vu} = \vec{0}^T.
    \label{eq:true_adjoint}
\end{equation}

Plugging this form of $\vlam$ into Eq. (\ref{eq:lagrange_deriv2}) gives the following expression for the gradient

\begin{equation}
    \frac{d\Lcal}{d\vphi} = \frac{d F}{d\vphi} = \int_0^T dt~  \Bigg[ \vlam^T(t) \frac{\partial \vg}{\partial \vphi}(t) \Bigg]
    \label{eq:lagrange_deriv3}.
\end{equation}

In practice, rather than solving Eq. (\ref{eq:true_adjoint}) backwards in time from $\vlam(T) = \vec{0}$ to $\vlam(0)$, we may instead define the time-reversed Lagrange multipliers $\tilde{\vlam}(t) \equiv \vlam(T - t)$ which have the initial condition $\tilde{\vlam}(0) = \vec{0}$ and may be solved forward in time.  After substituting $\tilde{\vlam}$ into Eq. (\ref{eq:true_adjoint}), and applying a transpose, the time-reversed Lagrange multipliers are the solution to

\begin{equation}
     \frac{\partial \vg}{\partial \dot{\vu}}^T \dot{\tilde{\vlam}} + \left( \frac{\partial \vg}{\partial \vu}^T - \frac{d}{dt}\frac{\partial \vg}{\partial \dot{\vu}}^T \right)\tilde{\vlam} + \frac{\partial f}{\partial \vu}^T = 0
    \label{eq:TR_adjoint}.
\end{equation}

Importantly, we note that while $\tilde{\vlam}$ is evaluated at time $t$ in Eq. (\ref{eq:TR_adjoint}), all other terms are evaluated at time $T - t$ as they were not time reversed.  Finally, in terms of $\tilde{\vlam}$, the gradient is
\begin{equation}
    \frac{d F}{d\vphi} = \int_0^T dt~  \Bigg[ \tilde{\vlam}^T(T-t) \frac{\partial \vg}{\partial \vphi}(t) \Bigg]
    \label{eq:lagrange_deriv_TR},
\end{equation}
or, written in terms of the original adjoint solution
\begin{equation}
    \frac{d F}{d\vphi} = \int_0^T dt~  \Bigg[ \vlam^T(t) \frac{\partial \vg}{\partial \vphi}(t) \Bigg]
    \label{eq:lagrange_deriv_TR2},
\end{equation}

\subsection{Application to Maxwell's Equations}

We now evaluate Eq. (\ref{eq:lagrange_deriv_TR}) for a system obeying Maxwell's equations.  We take our unknown to be the set of electric and magnetic fields, $\vu(t) = [\vg(t),~\ve(t)]^T$, and may derive the following expression for the constraint $\vg(\dot{\vu}, \vu, \vphi, t)$
\begin{align}
\vg(\dot{\vu}, \vu, \vphi, t) &= A(\vphi) \cdot \dot{\vu}(t) + B \cdot \vu(t) + \vc(t) = \vec{0}\\
  &= 
    \begin{bmatrix}
        -\mu & 0 \\ 0 & \epsilon(\vphi)
    \end{bmatrix}
    \cdot
    \begin{bmatrix}
        \dot{\vg} \\ \dot{\ve}
    \end{bmatrix}
    +
    \begin{bmatrix}
        -\sigma_H & \nabla \times \\
         \nabla \times & -\sigma_E
    \end{bmatrix}
    \cdot
    \begin{bmatrix}
        \vg\\
        \ve
    \end{bmatrix}
    +
    \begin{bmatrix}
        \vec{0} \\
        \vj
    \end{bmatrix} = \vec{0}
\label{eq:matrix_form}
\end{align}

This is assuming a time-independent permittivity, permeability and conductivity, although the same analysis can be applied in the time-dependent case.  Note also that we have assumed that only the only $\vphi$ dependence appears in the permittivity distribution $\epsilon(\phi)$, although this can be generalized to other degrees of freedom, such as the current source, $J$, or conductivity distributions, $\sigma$, without complication.

The derivatives of $\vg(\dot{\vu}, \vu, \vphi, t)$ are needed in the adjoint and gradient equations of Eq. (\ref{eq:TR_adjoint}) and Eq. (\ref{eq:lagrange_deriv_TR}), which are given by

\begin{align}
\frac{\partial \vg}{\partial \dot{\vu}} &= A(\vphi) =  
    \begin{bmatrix}
        -\mu & 0 \\ 0 & \epsilon(\vphi)
    \end{bmatrix} \\
\frac{\partial \vg}{\partial \vu} &= B = 
    \begin{bmatrix}
        -\sigma_H & \nabla \times \\
        \nabla \times & -\sigma_E
    \end{bmatrix} \\
\frac{\partial \vg}{\partial \vphi} &= \frac{\partial A}{\partial \vphi} \dot{\vu} = 
    \begin{bmatrix}
            0 & 0 \\
            0 & \epsilon'
    \end{bmatrix} 
    \cdot
    \begin{bmatrix}
        \dot{H} \\ \dot{E}
    \end{bmatrix}
    =
    \begin{bmatrix}
        0 \\ \epsilon' \cdot \dot{E}
    \end{bmatrix}
\end{align}
where $\epsilon' \equiv \frac{\partial \epsilon}{\partial \vphi}$ defines how the permittivity distribution changes with a change of parameters $\vphi$.  With these, the expression for the time-reversed Lagrange multipliers $\tilde{\vlam}$ from Eq. (\ref{eq:TR_adjoint}) becomes
\begin{equation}
    A^T \cdot \dot{\tilde{\vlam}} + B^T \cdot \tilde{\vlam} + \frac{\partial f}{\partial \vu}^T = 0
    \label{eq:TR_adjoint2},
\end{equation}

To express these in terms of Maxwell's Equations, we define the electromagnetic adjoint fields $\vlam(t) \equiv [\vh_\textrm{adj},~\ve_\textrm{adj}]^T$ and their evaluations at $T-t$ given by $\tilde{\vlam}(t) \equiv [\tilde{\vg}_\textrm{adj}(t),~\tilde{\ve}_\textrm{adj}(t)]^T$.  The latter are given by the solution to Eq. (\ref{eq:TR_adjoint2}),  written with $A^T$ and $B^T$ from by Maxwell's equations as
\begin{align}
    \begin{bmatrix}
        -\mu^T & 0 \\ 0 & \epsilon(\vphi)^T
    \end{bmatrix}
    \cdot
    \begin{bmatrix}
        \dot{\tilde{\vg}}(\tau) \\ \dot{\tilde{\ve}}(\tau)
    \end{bmatrix}
    +
    \begin{bmatrix}
        -\sigma_H^T & \nabla \times \\
         \nabla \times & -\sigma_E^T
    \end{bmatrix}
    \cdot
    \begin{bmatrix}
        \tilde{\vg}(\tau) \\ \tilde{\ve}(\tau)
    \end{bmatrix}
    +
    \begin{bmatrix}
        \left( \partial f / \partial \vh\right)^T(T-\tau) \\
        \left( \partial f / \partial \ve\right)^T(T-\tau)
    \end{bmatrix} = \vec{0}
\end{align}

Finally, in terms of $\tilde{\vg}$ and $\tilde{\ve}$, the gradient can be expressed as 

\begin{align}
    \frac{d F}{d \vphi} &= \int_0^T dt~  \Bigg[ \vlam(t)^T \frac{\partial \vg(t)}{\partial \vphi} \Bigg] \\
    &= \int_0^T dt~  \Bigg[ \tilde{\vlam}(T-t)^T \frac{\partial A}{\partial \vphi} \dot{\vu}(t) \Bigg] \\
    &= \int_0^T dt~  \Bigg[     
        \begin{bmatrix}
            \vh_\textrm{adj}(t), &
            \ve_\textrm{adj}(t)
        \end{bmatrix}^T
        \cdot
        \begin{bmatrix}
            0 \\
            \epsilon' \cdot \dot{\ve}(t)
        \end{bmatrix}
        \Bigg] \\
    &= \int_0^T dt~  \Bigg[ \ve_\textrm{adj}(t)  \cdot  \epsilon' \cdot \dot{\ve}(t) \Bigg] \\
    \label{eq:lagrange_deriv3}.
\end{align}

Practically, when solving these equations numerically, it can be more convenient to put this in a slightly different form.  By integrating by parts, one can move the time derivative to the adjoint field

\begin{align}
    \frac{dF}{d\vphi} &= \int_0^T dt~  \Bigg[ \ve_\textrm{adj}(t)  \cdot  \epsilon' \cdot \dot{\ve}(t) \Bigg] \\
    &= -\int_0^T dt~  \Bigg[ \dot{\ve}_\textrm{adj}(t)  \cdot  \epsilon' \cdot \ve(t) \Bigg] - \ve_\textrm{adj}(t)  \cdot  \epsilon' \cdot \ve(t) \Bigg|^T_0 \\
    &= -\int_0^T dt~  \Bigg[ \dot{\ve}_\textrm{adj}(t)  \cdot  \epsilon' \cdot \ve(t) \Bigg] - \cancel{ \ve_\textrm{adj}(T) }  \cdot  \epsilon' \cdot \ve(T) - \ve_\textrm{adj}(0)  \cdot  \epsilon' \cdot \cancel{ \ve(0) }  \\
    &= -\int_0^T dt~  \Bigg[ \dot{\ve}_\textrm{adj}(t)  \cdot  \epsilon' \cdot \ve(t) \Bigg]
    \label{eq:lagrange_deriv3}.
\end{align}
where the last line comes from the boundary conditions of $\ve(0) = \ve_\textrm{adj}(T) = \vec{0}$.

It is straightforward to show that $\dot{\ve}_\textrm{adj}(t)$ can be identified as the electric fields (no time derivative) found when running the adjoint problem with source $\frac{d}{dt} \frac{\partial f}{\partial \vu}$ instead of $\frac{\partial f}{\partial \vu}$, which is more practical when dealing with the algorithm where computing time derivatives requires careful treatment of finite-difference derivatives.

\section{Efficient Computation of the Adjoint Integral\label{appx:efficient}}

To compute the adjoint sensitivity, one must compute Eq. (\ref{eq:final_grad_adj}), copied below as
\begin{equation}
    \frac{dF_i}{d\vphi} =  \int_0^T dt~ \vlam_i(t)^T \cdot \frac{\partial \mA}{\partial \vphi} \cdot \vu(t)
    \label{eq:final_grad_adj_appx},
\end{equation}
where $\vlam$ is the adjoint field, $\vu(t)$ is the forward field and $\frac{\partial \mA}{\partial \vphi}$ is a rank three tensor of dimension ($N \times m \times N$) and $N$ is the number of grid points in the domain.

Computed naively, this integral would amount to a time complexity of $\Ocal(N^2 T m)$ and the computation of the full Jacobian would be $\Ocal(N^2 T m n)$.
However, in practice, we may eliminate the $m$-dependence of this integral, which results in the standard constant scaling of the adjoint method with respect to the number of inputs.

To do this, we first write Eq. (\ref{eq:final_grad_adj_appx}) in index notation, giving
\begin{equation}
    \left(\frac{dF}{d\vphi}\right)_k =  \int_0^T dt~ \sum_{i,j} \lambda_i \frac{\partial A_{ij}}{\partial \phi_k} u_j
    \label{eq:final_grad_adj_appx}.
\end{equation}

We note from Eq. (\ref{eq:maxwell}) that $\mA$ is diagonal, and therefore we may make the substitution
\begin{equation}
    \frac{\partial A_{ij}}{\partial \phi_k} = \delta_{ij} \frac{da_i}{d\phi_k},
\end{equation}
where $a_i$ is the value of $A_{ii}$, corresponding to either a relative permittivity or permeability.
Substituting into Eq. (\ref{eq:final_grad_adj_appx}) gives
\begin{align}
    \left(\frac{dF}{d\vphi}\right)_k &=  \int_0^T dt~ \sum_{i,j} \lambda_i \frac{\partial A_{ij}}{\partial \phi_k} u_j \\
     &=  \int_0^T dt~ \sum_{i,j} \lambda_i \delta_{ij} \frac{da_i}{d\phi_k} u_j \\
     &=  \int_0^T dt~ \sum_{i} \frac{da_i}{d\phi_k} \lambda_i u_i \\
     &=  \int_0^T dt~ \frac{d\vec{a}}{d\vphi}^T \cdot \left( \vlam(t) \odot \vu(t) \right) \\
     &=  \frac{d\vec{a}}{d\vphi}^T \cdot \left[ \int_0^T dt~ \left( \vlam(t) \odot \vu(t) \right) \right]
    \label{eq:final_grad_adj_better_appx},
\end{align}
where $\odot$ is element-wise vector multiplication and $\vec{a}$ refers to the vector along the diagonal of $\mA$.
Thus, we may first perform $\int_0^T dt~ \vlam(t) \odot \vu(t)$ in $\Ocal(NT)$ time.
In general, the matrix $\frac{d\vec{a}}{d\vphi}^T$ is sparse because each free parameter $\phi_k$ will only affect some subset of the spatial domain.
This means that the complexity of the multiplication in Eq. (\ref{eq:final_grad_adj_better_appx}) will be between $\Ocal(NT)$ and $\Ocal(NTm)$, but closer to $\Ocal(NT)$ in general.
Thus, the complexity of the total adjoint calculation is $\Ocal(NTn)$, as expected.

\section{Scaling Comparison of Methods\label{appx:scaling}}

We now examine how the speed and memory scaling of each of these methods compare methods compare as a function of the simulation parameters.  
As before, we assume a function $F$  with $m$ input parameters and $n$ output parameters.
The evaluation of $F$ involves running an FDTD simulation containing $N$ points in the spatial grid and $T$ time steps.  
In each case, the storage of Jacobian itself requires a memory storage of $\Ocal(mn)$ and a single FDTD run may be completed in $\Ocal(NT)$ time complexity and requires a memory storage of $\Ocal(N)$.

In the finite-difference approach, one must perform one independent FDTD simulation per input parameter.  
This leads to a time complexity of $\Ocal(NTm)$.
The finite-difference calculation itself requires storage of a column of the Jacobian and the subtraction of two vectors of length $m$, which adds no additional overhead in the memory complexity, which scales as $\Ocal(N + mn)$.

In FMD, one must first compute and store the full $\vu(t)$ sequence.  
Then, for each input parameter, one FDTD simulation must be performed along with the integral in Eq. (\ref{eq:forward_diff_start}), which gives a time complexity of $\Ocal(NT(n+m))$.
The corresponding memory complexity is $\Ocal(NT + mn)$

For the adjoint method, one must also compute and store the full $\vu(t)$ sequence.
Then, for each \textit{output} parameter, an adjoint FDTD simulation must be performed and the integral in Eq. (\ref{eq:final_grad_adj}) must be computed.

For the adjoint case, without any special techniques applied, one must store the entire forward field solution over time.  This results in a memory cost that scales as $\Ocal(NT)$.  Although the final gradient computation requires knowledge of the tensor, $\epsilon'$, which is of size $N \times P \times N$, there is no explicit need to store the full tensor in memory, so we ignore this contribution.

By contrast, the numerical derivative only requires storage of the electric fields at each time step, which is already required for the FDTD simulation, as well as the final figures of merit for each design parameter.  This results in a memory storage of $\Ocal(N) + \Ocal(P)$.

Like the numerical derivative, the forward difference approach requires running $P$ additional FDTD simulations.  However, in this case, one must use the forward field solution to construct a source for each of these simulations.  Therefore, in a naive implementation, one must run each of these simulations in parallel with the forward simulation, resulting in a memory storage of $\Ocal(NP)$.  However, this memory storage is independent of $T$, which may result in significant improvements in memory requirements in the case that the number of parameters is far fewer than the number of time steps.

\subsection{Time scaling}
We now examine the computational time complexity required for each method. The calculation of $u(t)$ requires a full FDTD simulation, which has a time complexity of $\Ocal(NT)$.  The first step of the adjoint gradient requires running two FDTD simulations to compute the forward and adjoint fields.  Therefore, this step has a time complexity of $\Ocal(NT)$.  The integral needed to compute the full sensitivity of Eq. (\ref{eq:final_grad_adj}), while generally requiring $\Ocal(NTP)$ operations, can be neglected in almost all practical considerations, as discussed in the \app Material.  Therefore, the time complexity of the adjoint gradient is $\Ocal(NT)$.

To compute the derivative using finite differences, one must run $P$ independent FDTD simulations then compute the finite difference derivative of numerical derivative expression from Eq. (\ref{eq:finite_diff}).  This gives a time complexity of $\Ocal(NTP)$.

The forward difference approach also requires running $P$ independent simulations.  To compute the gradient, one may integrate Eq. (\ref{eq:forward_diff_start}), within the FDTD loop, which does not add additional scaling to the time complexity.  Therefore, like numerical derivatives, the forward differentiation method has a time complexity of $\Ocal(NTP)$.

\subsection{Summary}

A summary of the time and memory requirements of each method are given in Table \ref{tb:table_order}.

\begin{table}[]
\begin{tabular}{|l|l|l|}
\hline
\textbf{Method} & \textbf{Time Complexity}  & \textbf{Memory Complexity}   \\ \hline
Numerical     & $\Ocal(NTm)$   & $\Ocal(N)$ \\ \hline
FMD & $\Ocal(NTm)$ & $\Ocal(NT+Nn)$ \\ \hline
Adjoint & $\Ocal(NTn)$ & $\Ocal(NT+Nm)$  \\ \hline
\end{tabular}
\caption{Comparison of memory and speed complexity of the three different gradient computation methods examined in this work.  $N$ is the number of grid cells.  $T$ is the number of time steps. $m$ and $n$ are the number of input and output variables of the function $F$.  We assume $n$ and $m$ are each less than or equal to $N$.}
\label{tb:table_order}
\end{table}

\end{document}